\documentclass{article}
\usepackage[margin=1.5cm]{geometry}
\usepackage{graphicx} 
\usepackage[utf8]{inputenc}
\usepackage{longtable}
\usepackage{url}
\usepackage{algorithm}
\usepackage{algpseudocode}
\usepackage{amsmath} 
\usepackage{array}
\usepackage{authblk}
\usepackage{hyperref}

\title{Bhakti: A Lightweight Vector Database Management System for Endowing Large Language Models with Semantic Search Capabilities and Memory}

\author[1]{Zihao Wu \thanks{Email: vortez.wohl@gmail.com}}

\affil[1]{Fujian University of Technology}

\date{\today}

\begin{document}

\maketitle

\begin{abstract}
    \noindent
    With the rapid development of big data and artificial intelligence technologies, the demand for effective processing and retrieval of vector data is growing. Against this backdrop, I have developed the Bhakti vector database \footnote{The code is available at: \url{https://github.com/vortezwohl/Bhakti}}, aiming to provide a lightweight and easy-to-deploy solution to meet the storage and semantic search needs of small and medium-sized datasets. Bhakti  supports a variety of similarity calculation methods and a domain-specific language (DSL) for document-based pattern matching pre-filtering, facilitating migration of data with its portable data files, flexible data management and seamless integration with Python3. Furthermore, I propose a memory-enhanced large language model dialogue solution based on the Bhakti database, which can assign different weights to the question and answer in dialogue history, achieving fine-grained control over the semantic importance of each segment in a single dialogue history. Through experimental validation, my method shows significant performance in the application of semantic search and question-answering systems. Although there are limitations in processing large datasets, such as not supporting approximate calculation methods like HNSW, the lightweight nature of Bhakti gives it a clear advantage in scenarios involving small and medium-sized datasets. \\
    \textbf{Keywords:} Vector Database, Semantic Search, RAG, Conversational AI
\end{abstract}

\section{Introduction}

The advent of big data and the remarkable advancements in artificial intelligence have ushered in an era where the efficient processing and retrieval of vector data are of paramount importance. Vector databases, specifically designed to handle vector data, offer efficient similarity search and data management functionalities, making them indispensable in various applications such as semantic retrieval, recommendation systems, and conversational AI \cite{Yang2023}. The core technologies of vector databases encompass vector space modeling, distance metrics, index construction, and optimized retrieval algorithms \cite{Bowman2023}. With the breakthroughs in large language models (LLMs), the interest in vector databases has been escalating, prompting the development of Bhakti, a vector database system that caters to the market's demand for lightweight, availability, and ease of use.

Bhakti is a lightweight vector database tailored for the storage and retrieval of medium-scale datasets. It supports K-nearest neighbor semantic search \cite{kramer2013k} and a domain-specific language (DSL) for document-based pattern matching pre-filtering, facilitating migration of data with its portable data files and seamless integration with Python3. Bhakti stands out with its support for various similarity calculations, although it currently focuses on precise computations rather than approximate methods like HNSW \cite{Bentley1975}. This database is not designed to compete on performance metrics with more robust vector database solutions but rather to provide a user-friendly, deployable, and adaptable system for scenarios where ease of use and data portability are prioritized.

The integration of Bhakti with large language models enhances their semantic search capabilities, effectively granting them a form of memory. This memory augmentation is achieved through a novel approach that assigns varying weights to the query and answer pairs within human-machine dialogues. Furthermore, it allows for the fine-tuning of the semantic significance of specific phrases within the dialogue history. This method of dialogue history vectorization embeds the nuances of each conversation segment, enabling more granular control over the importance of individual text segments \cite{Asai2023}.

When compared to other vector databases, Bhakti's lightweight nature positions it as an ideal solution for environments with limited resources or specific requirements for ease of deployment and data mobility. While it may not match the performance of databases that support approximate nearest neighbor search, its precision and simplicity make it a strong contender for applications where these features are more critical.

In the realm of document embedding, Bhakti's approach diverges from traditional RAG (Retrieval-Augmented Generation) \cite{Huang2024SurveyRetrievalAugmentedTextGeneration} schemes by introducing weighted query-answer pairs and nuanced control over dialogue history. This innovation allows for a more dynamic and context-aware interaction, particularly beneficial in long-term conversational contexts.

The development of Bhakti and its integration with large language models involved showcasing its capabilities, such as multiple semantic similarity support and flexible data management, and demonstrating its performance with medium-scale datasets. The research also focused on the development of a memory-augmented dialogue system that leverages Bhakti's semantic search capabilities. This system not only enhances the dialogue's coherence and consistency but also significantly improves the chatbot's memorability, engagingness, and humanness.

The aim of this study is to develop the Bhakti vector database to meet the storage and semantic search needs of small and medium-sized datasets. The specific research objectives are as follows:

\begin{enumerate}
    \item To implement a vector database that supports multiple similarity calculation methods to meet the processing and retrieval of vector data in various scenarios.
    
    \item To provide seamless integration with Python3, enhancing the flexibility of data management.
    
    \item To propose a memory-enhanced large language model dialogue solution based on the Bhakti database, which can assign different weights to questions and answers in dialogue history, achieving fine-grained control over the semantic importance of each segment in a single dialogue history.
    
    \item To demonstrate the application performance of this method in semantic search and question-answering systems through experimental validation.
    
    \item To analyze the limitations of Bhakti in processing large datasets and explore its advantages in scenarios involving small and medium-sized datasets.
\end{enumerate}

\section{Related Work}

In the domain of information retrieval, vector databases play a pivotal role in enhancing the capabilities of search engines. By converting text queries into vector embeddings, search engines can use K-nearest neighbor search algorithms to identify relevant documents more effectively. This semantic search approach goes beyond mere keyword matching, providing users with more meaningful results that align with their queries' context and intent \cite{Yang2023}.

Vector databases also underpin recommendation systems by enabling the representation of items as vectors. Through K-nearest neighbor search, these systems can identify items similar to those a user has previously engaged with, offering personalized recommendations. This approach is particularly effective in dynamic environments where user preferences are continually evolving \cite{Bowman2023}.

The brute force method for K-nearest neighbor search, despite its simplicity, serves as a foundational approach against which the efficiency of more sophisticated methods can be measured. It involves a straightforward computation of distances from the query point to all other points in the dataset, followed by the selection of the 'k' closest points. While this method is conceptually clear and easy to implement, its scalability is limited, making it unsuitable for applications involving large-scale data \cite{Bentley1975}.

The integration of vector databases with LLMs opens up a plethora of possibilities in various domains. In semantic search, for instance, vector databases can support search across different types of data, such as text, images, audio, or video, enhancing the user experience by providing more diverse and rich results \cite{Asai2023}. Furthermore, vector databases can enable real-time knowledge search, allowing LLMs to access the most up-to-date and accurate information from various sources, thereby improving the user’s awareness and understanding \cite{Liu2023}.

In recommendation systems, vector databases can power the identification of similar or complementary items based on their vector representations. This capability allows LLMs to provide explanations for recommendations and offer additional information or reviews, enhancing user satisfaction \cite{Whitehouse2023}.

The potential applications of vector databases in conjunction with LLMs are vast and varied. They range from providing long-term memory for LLMs by storing relevant documents or information in vector form to enabling semantic search and powering recommendation systems. These applications highlight the synergy between vector databases and LLMs, showcasing how they can work together to create more intelligent and interactive systems \cite{Huang2023}.

\section{System Design and Implementation}

\subsection{System Architecture}

\begin{figure}[h]
    \centering
    \includegraphics[width=0.4\textwidth]{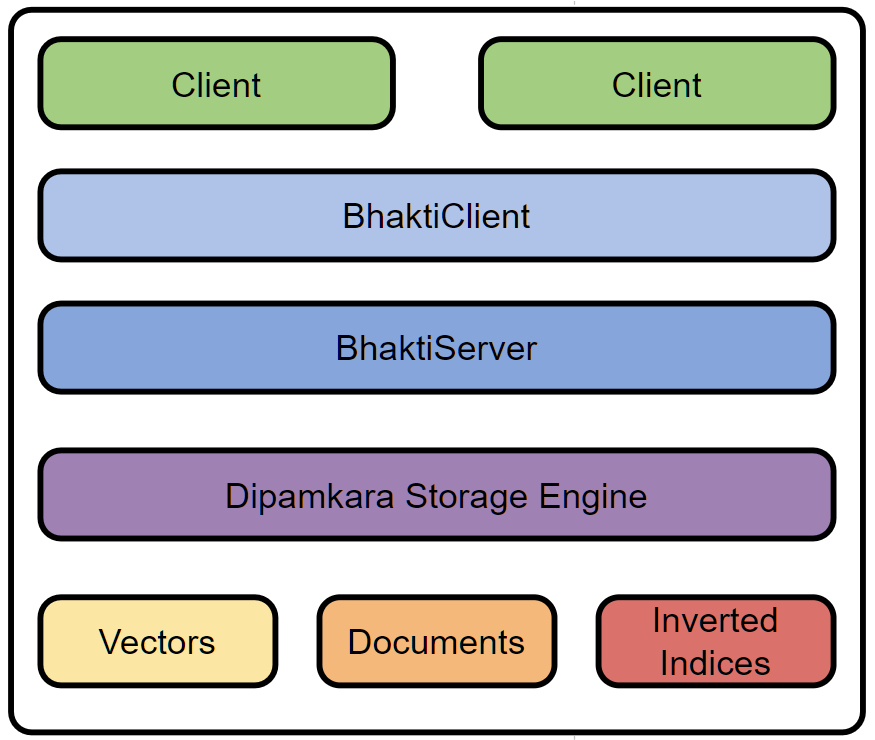}
    \caption{High-Level Architecture}
    \label{fig:image_1}
\end{figure}

The high-level architecture of the Bhakti system, as depicted in Figure \ref{fig:image_1}, is a client-server model that revolves around the BhaktiClient and BhaktiServer, with the core functionality provided by the Dipamkara Storage Engine which is the default Storage Engine for Bhakti. The BhaktiClient serves as the interface for users to interact with the system, allowing them to perform operations such as data insertion, querying, and management of vector data. It is designed to be user-friendly and provides a reactive approach to handling database operations.

The BhaktiServer acts as the intermediary between the client and the storage engine. It is responsible for processing client requests, managing network communication, and coordinating the execution of database operations. The server is built to be highly responsive and efficient, ensuring that the system can handle a large volume of data with minimal latency.

At the heart of the Bhakti system lies the Dipamkara Storage Engine, which is tasked with the management of documents, inverted indices, and vector indices. This engine is designed to be thread-safe and supports asynchronous operations, making it suitable for handling concurrent database transactions in a multi-threaded environment. The engine utilizes advanced indexing techniques, such as inverted indices, to enhance the speed and accuracy of similarity searches.

The inverted indices in the Dipamkara Storage Engine play a pivotal role in the system's performance. They facilitate the pre-filtering of documents based on content before vector similarity search, enabling the system to efficiently narrow down relevant documents. This feature is particularly beneficial for applications that demand swift and precise data retrieval, such as recommendation systems and semantic search engines.

The architecture is designed with scalability and performance in mind. The separation of concerns between the client, server, and storage engine allows for independent scaling and optimization of each component. This modular approach also facilitates easier maintenance and upgrades, as each part of the system can be improved without affecting the others.


\begin{figure}[h]
    \centering
    \includegraphics[width=0.6\textwidth]{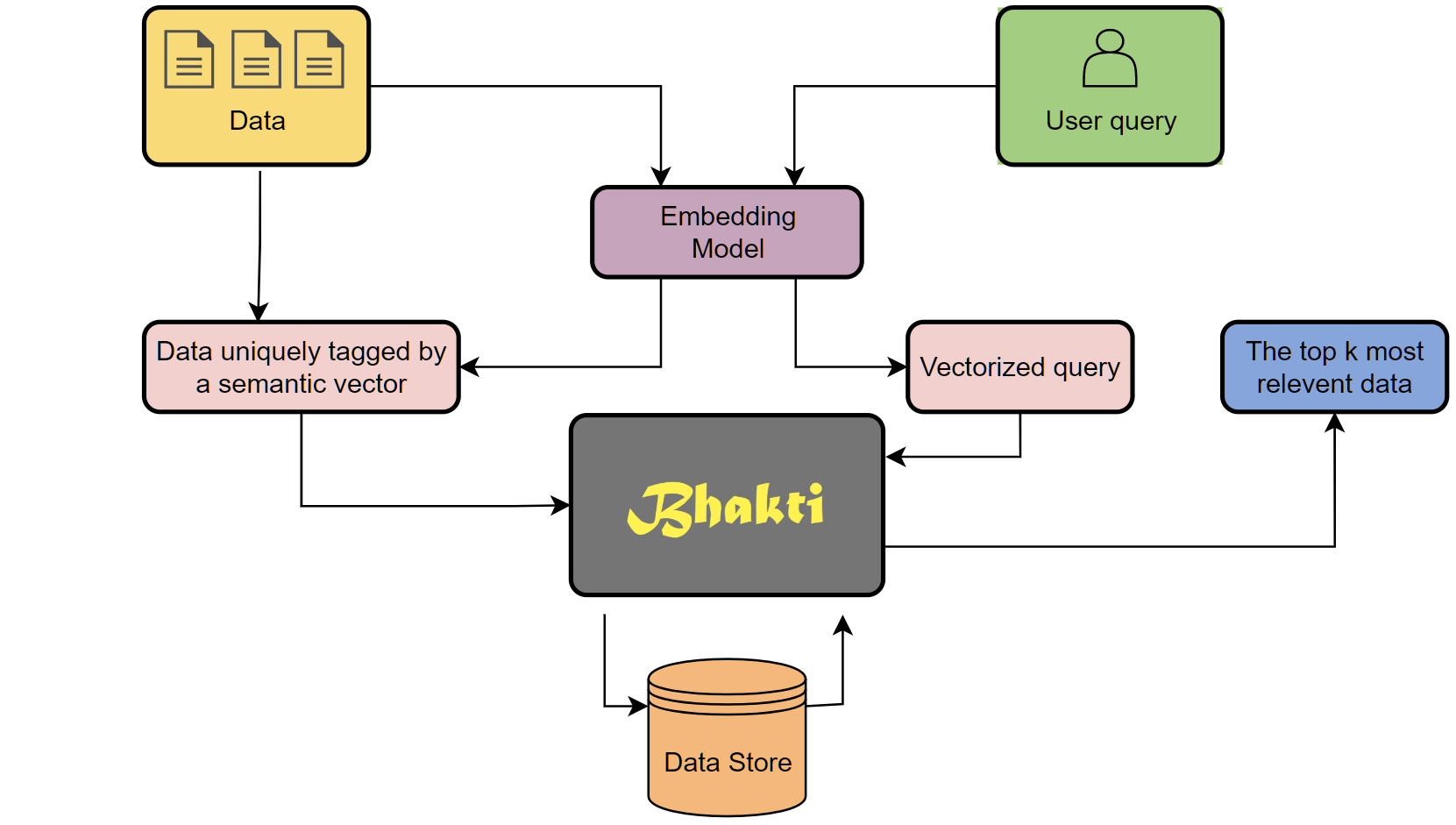}
    \caption{Semantic Search Process}
    \label{fig:image_2}
\end{figure}

The Semantic Search Process, as depicted in Figure \ref{fig:image_2}, is an intricate system designed to facilitate the retrieval of data based on semantic similarity rather than literal matching. This system is particularly adept at understanding the context and nuances of user queries, thereby providing more relevant and accurate search results.

The interaction model of the semantic search process can be elucidated through the following sequential steps:

\paragraph{User Query} The process commences with the user submitting a query to the system. This query is not merely a set of keywords but a comprehensive input that the system aims to understand in its entirety.
    
\paragraph{Embedding} Upon receiving the query, the system translates it into a semantic vector \cite{Moka-Massive-Mixed-Embedding}, an embedding that encapsulates the essence of the query. This transformation is critical as it allows the system to interpret the query in a manner that can be effectively used for comparison with the data stored.

\paragraph{Data Store} The Data Store plays a dual role in this process. Firstly, it serves as the repository where the vectorized data is stored, each uniquely tagged by a semantic vector. This tagging is a preprocessing step that enables efficient comparison with the vectorized query. Secondly, the Data Store integrates the user's input with the system's output, embedding them into a unified representation that is then stored in the database. This dual function allows the system not only to retrieve data but also to learn from each interaction, refining its understanding and improving future search accuracy.
    
\paragraph{Retrieval} The system then performs a comparison between the query vector and the vectors representing the data in the store. It calculates the semantic similarity to identify the top k most relevant data.
    
\paragraph{Output} The system presents the user with the top k most relevant data that closely matches the semantic vector of the query. This output is the culmination of the semantic search process, providing the user with data that is not only relevant but also contextually significant. \\

\noindent
This interaction model is a testament to the power of semantic search in understanding the underlying meaning of user queries and its ability to deliver data that is contextually rich and meaningful. It shifts the paradigm from keyword matching to a more sophisticated level of interaction that is akin to human understanding, bridging the gap between user intent and data retrieval.

\subsection{Network Communication}

\subsubsection{Pipeline Architecture}

The pipeline architecture consists of a series of stages, each responsible for a specific task in the data processing workflow. Data flows through these stages, with each stage transforming the data in some way before passing it on to the next stage. The key characteristics of this architecture include:

\paragraph{Modularity} Each stage of the pipeline is an independent module, performing a distinct function such as data ingestion, processing, analysis, and output generation.
    
\paragraph{Sequential Processing} Data moves through the pipeline in a sequential order, ensuring that each stage receives the data in a timely manner and processes it before passing it to the next stage.
    
\paragraph{Asynchronous Execution} The pipeline supports asynchronous execution, allowing for non-blocking data processing and improving overall throughput.
    
\paragraph{Error Handling} The architecture includes robust error handling mechanisms, ensuring that errors in one stage do not halt the entire pipeline.
    
\paragraph{Extensibility} New stages can be easily added to the pipeline to accommodate additional processing requirements or to scale the system. \\

\noindent
I implement a basic pipeline architecture with Python's asyncio library for asynchronous programming. The \texttt{PipelineStage} class serves as the base class for each stage in the pipeline, while the \texttt{Pipeline} class manages the execution of the pipeline.

\paragraph{PipelineStage Class} This abstract class defines the interface for each pipeline stage. It must be subclassed to implement the \texttt{do} method, which performs the actual data processing.

\paragraph{Pipeline Class} This class initializes the pipeline with a queue of stages, input/output context, end-of-file marker, and additional context. The \texttt{launch} method orchestrates the execution of the pipeline, iterating over each stage and passing data between them.

\subsubsection{Request-Response Communication Model}

The Request-Response communication model \cite{CSDN_http} is a fundamental concept in computer networking and forms the basis of client-server interactions. This model is characterized by a client sending a request to a server, which then processes the request and returns a response back to the client.

The process begins with a client, initiating a communication by sending a request to a server. This request contains all the necessary information the server needs to understand and fulfill the client's needs. The request follows a specific format defined by the communication protocol being used.

Upon receiving the request, the server parses the information, executes the appropriate actions, such as retrieving data from a database, processing a transaction, or serving a web page, and then constructs a response. The response includes a status code indicating the outcome of the request, headers that provide additional information about the response, and a body containing the requested data or resources.

The client then receives the response and processes it according to its capabilities. For example, a web browser might render an HTML page \cite{CSDN_http}, while an email client might display a message. The client's software is responsible for interpreting the response and presenting it to the user in a meaningful way.

\noindent
The Request-Response model has several key characteristics that define its behavior:

\paragraph{Synchronous Interaction} The client sends a request and then waits for a response. This synchronous interaction can provide more reliable interaction because it ensures the order and timeliness of requests and responses. Synchronous interaction also allows the client to handle errors or exceptions immediately. This immediate feedback enables quicker error handling and recovery, reducing the likelihood of cascading failures.
    
\paragraph{Statelessness} In the context of stateless protocols like HTTP, each request from the client contains all the information the server needs to process the request. The server does not retain any information about previous requests from the same client.
    
\paragraph{Modularity} The model promotes modularity by separating the concerns of the client and the server. The client is responsible for the user interface and experience, while the server manages data processing and storage.
    
\paragraph{Scalability} Servers can be scaled to handle a large number of requests by distributing the load across multiple servers or by using load balancers. \\

\noindent
The communication protocol is structured around JSON (JavaScript Object Notation) strings \cite{crockford_json}, which are human-readable and easy to parse by machines. The protocol consists of two main parts: the request and the response. The use of JSON for my communication protocol offers several advantages:

\paragraph{Simplicity} JSON is lightweight and easy to implement, making it accessible for various client and server platforms.
    
\paragraph{Extensibility} The protocol can be easily extended with additional parameters or commands without breaking compatibility.
    
\paragraph{Interoperability} JSON is widely supported, allowing for seamless integration with other systems and languages. \\

\noindent
A request in this protocol is formatted as follows:

\begin{longtable}{p{0.8\textwidth}}
\{
\\
\ \ "db\_engine": "<engine\_value>",\\
\ \ "opt": "<operation\_type>",\\
\ \ "cmd": "<command>",\\
\ \ "param": \{
\\
\ \ \ \ "<param\_name1>": "<param\_value1>",\\
\ \ \ \ "<param\_name2>": "<param\_value2>",\\
\ \ \ \ \ldots\\
\ \ \ \}\\
\}
\end{longtable}

\noindent
For example, a request to create an index might look like this:

\begin{longtable}{p{0.8\textwidth}}
\{
\\
\ \ "db\_engine": "dipamkara",\\
\ \ "opt": "create",\\
\ \ "cmd": "create\_index",\\
\ \ "param": \{
\\
\ \ \ \  "index": "my\_index",\\
\ \ \ \ "detailed": false\\
\ \ \ \}\\
\}
\end{longtable}

\noindent
The response format is designed to provide clear feedback to the client regarding the outcome of the request:

\begin{longtable}{p{0.8\textwidth}}
\{
\\
\ \  "state": "<state>",\\
\ \ "message": "<message>",\\
\ \ "data": "<data>"\\
\}
\end{longtable}

\noindent
A successful response might be:

\begin{longtable}{p{0.8\textwidth}}
\{
\\
\ \  "state": "OK",\\
\ \ "message": "",\\
\ \ "data": "true"\\
\}
\end{longtable}

\noindent
Or an exception response might be:

\begin{longtable}{p{0.8\textwidth}}
\{
\\
\ \  "state": "Exception",\\
\ \ "message": "Read timeout",\\
\ \ "data": null\\
\}
\end{longtable}

\noindent
The implementation of the request-response communication involves the following steps:

\paragraph{Client Request} The client constructs a JSON string according to the request format, specifying the database engine, operation type, command, and parameters. This string is then sent to the server, depending on my network protocol.
    
\paragraph{Server Processing} Upon receiving the request, the server parses the JSON string to extract the operation details. The server then performs the specified database operation using the provided parameters.
    
\paragraph{Server Response} After completing the operation, the server constructs a JSON string according to the response format. This includes the state of the operation (e.g., "OK" or "Exception"), an optional message providing more details, and the data resulting from the operation.
    
\paragraph{Client Reception} The client receives the JSON response and parses it to determine the outcome of the request. The client's application logic then handles the response, whether it involves updating the database, logging an error, or taking further actions based on the response data.

\subsection{Storage Engine}

Dipamkara is my storage engine which is sophisticated, thread-safe and designed for managing documents and performing vector indexing based on NumPy \cite{harris2020array}. It incorporates a custom DSL (Domain-Specific Language) for querying, making it a versatile tool for various data management tasks. This paper delves into the data model of the Dipamkara storage engine, exploring its architecture, data structures, and operations.

\subsubsection{Architecture}

Dipamkara's architecture is centered around several core components that facilitate data storage, retrieval, and manipulation. These components include:

\begin{itemize}
    \item A document storage system that caches documents for rapid access.
    \item An inverted index for efficient querying of documents based on specific keys.
    \item A vector storage system that maps vectors to document IDs, enabling fast vector-based searches.
    \item An auto-increment mechanism for assigning unique IDs to new documents.
    \item Asynchronous saving capabilities to ensure data persistence.
\end{itemize}

\subsubsection{Data Structure}

The engine utilizes several key data structures to manage data:

\paragraph{Vector Index} Vectors are stored as strings in a dictionary that maps unique vector representations to document IDs. This enables efficient retrieval of documents based on vector similarity.

\paragraph{Inverted Index} The inverted index is a dictionary that maps keys to a dictionary of vectors and their corresponding document IDs. This structure allows for quick lookup of documents based on specific attributes.

\paragraph{Document Storage} Documents are stored in a dictionary that maps document IDs to their respective documents. This structure supports fast access and modification of documents.

\subsubsection{Operations}

\paragraph{Creation} The \texttt{create} method allows for the creation of new documents associated with specific vectors. It also supports the creation of indices for efficient querying.
    
\paragraph{Retrieval} Documents can be retrieved using various methods, including vector queries and DSL-based vector queries. These methods enable the retrieval of documents based on vector similarity and specific criteria.
    
\paragraph{Modification} The engine supports the modification of documents through methods like \texttt{mod\_doc\_by\_vector}, which allows for updates to specific fields within a document.
    
\paragraph{Deletion} Documents can be deleted using vector-based removal or DSL-based queries, enabling the removal of documents based on specific criteria.
    
\paragraph{Concurrency Control} To ensure thread safety, Dipamkara employs a locking mechanism that applies to critical sections of the code. This mechanism prevents data corruption and ensures consistent operations.
    
\paragraph{Persistence and Caching} Dipamkara supports asynchronous saving of data to files, ensuring data persistence across system restarts. Additionally, the engine can be configured to cache documents in memory for faster access.

\subsubsection{Similarity Metrics}

\noindent\textbf{Cosine Distance:} For two vectors $\mathbf{A}$ and $\mathbf{B}$, the Cosine distance is calculated as:

\begin{equation}
d_{COS}(\mathbf{A}, \mathbf{B}) = 1-\cos(\theta) = 1-\frac{\mathbf{A} \cdot \mathbf{B}}{\|\mathbf{A}\| \|\mathbf{B}\|}
\end{equation}

Where:

\begin{itemize}
    \item $\mathbf{A} \cdot \mathbf{B}$ is the dot product of vectors $\mathbf{A}$ and $\mathbf{B}$
    \item $\|\mathbf{A}\|$ and $\|\mathbf{B}\|$ are the magnitudes (Euclidean norms) of vectors $\mathbf{A}$ and $\mathbf{B}$ respectively
\end{itemize}

For $n$-dimensional vectors, this formula expands to:

\begin{equation}
d_{COS}(\mathbf{A}, \mathbf{B}) = 1-\frac{\sum_{i=1}^n A_i B_i}{\sqrt{\sum_{i=1}^n A_i^2} \sqrt{\sum_{i=1}^n B_i^2}}
\end{equation}

\noindent \textbf{Euclidean Distance:} For two points $\mathbf{P} = (p_1, p_2, \ldots, p_n)$ and $\mathbf{Q} = (q_1, q_2, \ldots, q_n)$ in $n$-dimensional Euclidean space, the Euclidean distance is calculated as:

\begin{equation}
d(\mathbf{P}, \mathbf{Q}) = \sqrt{\sum_{i=1}^n (p_i - q_i)^2}
\end{equation}

Where:

\begin{itemize}
    \item $\|\mathbf{P} - \mathbf{Q}\|$ denotes the Euclidean norm of the difference vector
    \item $\cdot$ represents the dot product
\end{itemize}

\noindent \textbf{L2-normalized Euclidean Distance:} For two vectors $\mathbf{A} = (a_1, a_2, \ldots, a_n)$ and $\mathbf{B} = (b_1, b_2, \ldots, b_n)$ in $n$-dimensional space, the L2-normalized Euclidean distance is calculated as follows:

First, we normalize the vectors:

\begin{equation}
\hat{\mathbf{A}} = \frac{\mathbf{A}}{\|\mathbf{A}\|_2} = \frac{(a_1, a_2, \ldots, a_n)}{\sqrt{\sum_{i=1}^n a_i^2}}
\end{equation}

\begin{equation}
\hat{\mathbf{B}} = \frac{\mathbf{B}}{\|\mathbf{B}\|_2} = \frac{(b_1, b_2, \ldots, b_n)}{\sqrt{\sum_{i=1}^n b_i^2}}
\end{equation}

Then, we calculate the Euclidean distance between these normalized vectors:

\begin{equation}
d_{L2N}(\mathbf{A}, \mathbf{B}) = \|\hat{\mathbf{A}} - \hat{\mathbf{B}}\|_2 = \sqrt{\sum_{i=1}^n (\hat{a_i} - \hat{b_i})^2}
\end{equation}

Where:

\begin{itemize}
    \item $\|\cdot\|_2$ denotes the L2 norm (Euclidean norm)
    \item $\hat{\mathbf{A}}$ and $\hat{\mathbf{B}}$ are the L2-normalized representations of $\mathbf{A}$ and $\mathbf{B}$
\end{itemize}

\noindent \textbf{Standardized Euclidean Distance:} For two points $\mathbf{P} = (p_1, p_2, \ldots, p_n)$ and $\mathbf{Q} = (q_1, q_2, \ldots, q_n)$ in $n$-dimensional space, the standardized Euclidean distance is defined as:

\begin{equation}
d_{SE}(\mathbf{P}, \mathbf{Q}) = \sqrt{\sum_{i=1}^n \frac{(p_i - q_i)^2}{\sigma_i^2}}
\end{equation}

Where:

\begin{itemize}
    \item $\sigma_i^2$ is the variance of the $i$-th feature across the entire dataset
\end{itemize}

This can also be expressed in terms of Z-scores:

\begin{equation}
d_{SE}(\mathbf{P}, \mathbf{Q}) = \sqrt{\sum_{i=1}^n (Z_{p_i} - Z_{q_i})^2}
\end{equation}

Where:

\begin{itemize}
    \item $Z_{p_i}$ and $Z_{q_i}$ are the Z-scores of $p_i$ and $q_i$ respectively: \begin{equation}Z_{p_i} = \frac{p_i - \mu_i}{\sigma_i}, \quad Z_{q_i} = \frac{q_i - \mu_i}{\sigma_i}\end{equation}
    \item $\mu_i$ is the mean of the $i$-th feature across the dataset.
\end{itemize}

\noindent \textbf{Chebyshev Distance:} For two points $\mathbf{P} = (p_1, p_2, \ldots, p_n)$ and $\mathbf{Q} = (q_1, q_2, \ldots, q_n)$ in $n$-dimensional space, the Chebyshev distance is defined as:

\begin{equation}
d_{\infty}(\mathbf{P}, \mathbf{Q}) = \max_{i} |p_i - q_i|
\end{equation}

This can also be expressed using the limit of the p-norm:

\begin{equation}
d_{\infty}(\mathbf{P}, \mathbf{Q}) = \lim_{p \to \infty} \left( \sum_{i=1}^n |p_i - q_i|^p \right)^{\frac{1}{p}}
\end{equation}

Where:
\begin{itemize}
    \item $\max$ denotes the maximum function
    \item $|\cdot|$ represents the absolute value
\end{itemize}


\subsubsection{DSL}

Dipamkara's DSL expression allows for pattern matching queries before similarity calculation. The DSL supports a range of operations, including logical and relational operators, enabling sophisticated data retrieval.



    
    

\subsection{Testing On Different Scale of Data}

The performance test was conducted by varying the scale of data stored in the Bhakti database, ranging from 1 to 5000 data entries. For each scale, the query time was measured in milliseconds (ms). The test was performed twice for each data scale: once without using an inverted index for pre-filtering (represented by the green line) and once with the inverted index enabled for pre-filtering (represented by the blue line). The purpose of this approach was to assess the impact of using an inverted index on query performance.

\begin{figure}[h]
    \centering
    \includegraphics[width=0.4\textwidth]{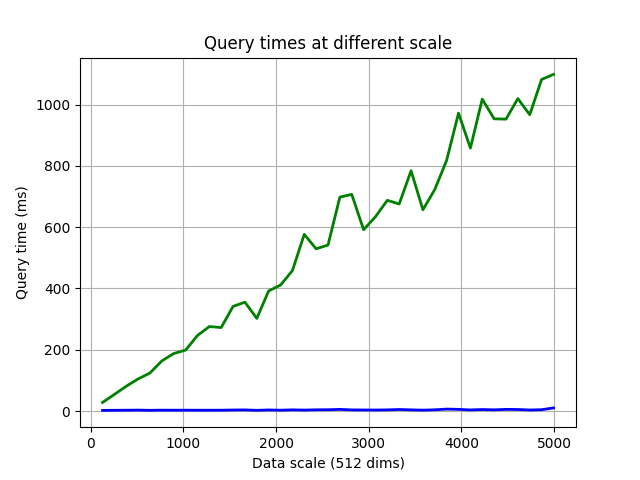}
    \caption{Query Time At Different Scale of Data}
    \label{fig:image_4}
\end{figure}

The results, as depicted in the provided Figure \ref{fig:image_4} show a clear trend in the query times for both scenarios. Initially, as the data scale increases from 1000 to 4000, the query times for both methods increase gradually. However, beyond 4000 data entries, the query times for both methods begin to rise more sharply.

A significant observation is the performance difference between the two methods. The green line, representing queries without the use of an inverted index, shows consistently higher query times compared to the blue line, which represents queries with an inverted index. This indicates that the use of an inverted index significantly improves the query performance of the Bhakti database.

The testing also reveals that at around 4000 data entries, the query times for both methods approach or exceed 1 second. This threshold is particularly noteworthy, as it suggests that the Bhakti database may not be the most suitable choice for scenarios involving very large datasets, where query times of 1 second or more could be considered impractical.

\section{Memory-based Dialogue Solution}

\begin{figure}[h]
    \centering
    \includegraphics[width=0.6\textwidth]{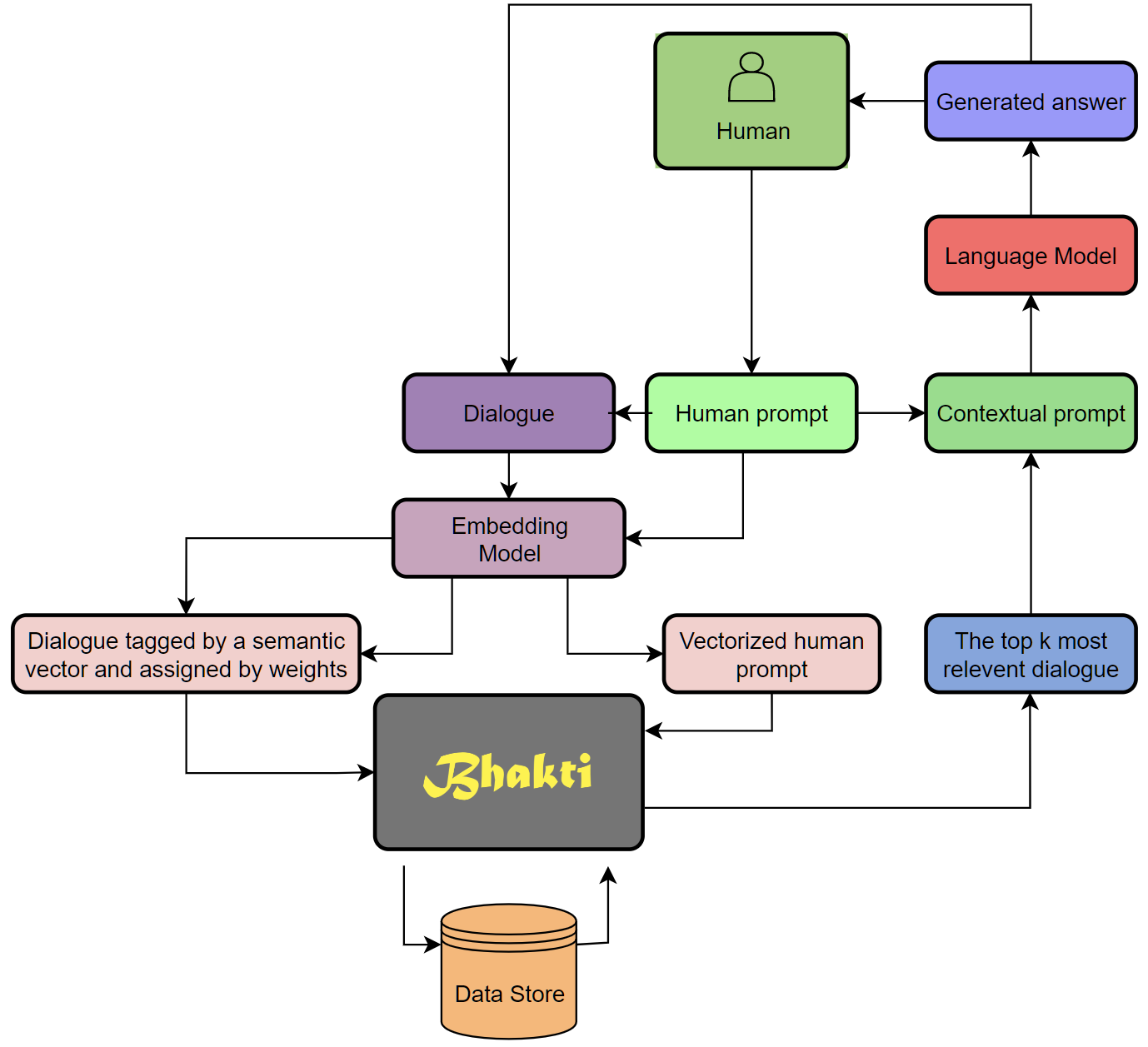}
    \caption{Memory-based Dialogue}
    \label{fig:image_3}
\end{figure}

As shown in Figure \ref{fig:image_3}, I presents an innovative Memory-based Dialogue Solution that revolutionizes the way dialogue systems interact with humans. By employing semantic vectorization and a dialogue data store, the system can effectively understand and respond to human prompts with high relevance and context awareness. The solution is designed to enhance user experience in various applications, including customer service, personal assistants, and more.

\subsection{Human-Machine Interface}

The system is designed to interact with humans through a natural language interface, allowing users to communicate their needs and expectations in a conversational manner.

\subsection{Language Model}

At the core of the system is a sophisticated language model \cite{qwen2.5} that processes the human prompts and generates contextually relevant responses.

\subsection{Embedding Model}

The embedding model \cite{Moka-Massive-Mixed-Embedding} is responsible for converting the dialogue tagged by semantic vectors into a vectorized form, which can be more effectively processed and compared.

\subsection{Memory Updating}

The Memory Updating process is a fundamental aspect of maintaining an effective dialogue system. It involves encoding new conversations into a vector space and storing them in a database for future retrieval.

\subsubsection{Dialogue Embedding}
The first step in the Memory Updating process is to embed the text of the user's query and the system's response into a dense vector. This is achieved using a pre-trained embedding model \cite{Moka-Massive-Mixed-Embedding}, which converts the text into a numerical representation that can be processed by the system. I propose an innovative approach of dialogue history embedding which is described in \ref{Dialogue History Vectorization}.

\subsubsection{Storing Dialogue}

Once the text has been embedded into a vector, the next step is to store these vectors along with the corresponding conversation data in a database.

The system interacts with vector database to store and retrieve conversation data. This involves creating a new record for each conversation, which includes the query vector, answer vector, user ID, bot ID, and timestamp.

The vectors are stored in the database in a way that allows for easily retrieval. This typically involves indexing the vectors based on their semantic similarity, enabling the system to quickly find relevant conversations when needed.

\subsection{Memory Retrieval}

The Memory Retrieval process is a fundamental aspect of maintaining an effective dialogue system. It involves searching the system's memory for relevant conversations based on the user's current query.

\subsubsection{Query Embedding}

The first step in the Memory Retrieval process is to encode the user's query into a vector space representation. The vectorization process involves converting the text of the user's query into a high-dimensional vector space, where semantically similar words or phrases are mapped to nearby points in the space.

\subsubsection{Recall}

Once the query has been embedded into a vector, the next step is to search the system's memory for relevant conversations. The retrieval algorithm is responsible for identifying the most relevant conversations based on the user's query. This typically involves calculating the similarity between the query vector and the vectors stored in the database, and ranking the results based on this similarity. The Memory Retrieval process returns a list of relevant conversations, which can be used to inform the system's response to the user's query. The retrieval results are ranked based on their similarity to the user's query. This ranking allows the system to prioritize the most relevant conversations when generating its response. The system uses the retrieved conversations to generate a response to the user's query. This may involve selecting the most relevant conversation and using it as the basis for the response, or combining information from multiple conversations to create a more comprehensive response. The Memory Retrieval process allows the system to provide personalized responses based on the user's past interactions. By leveraging the system's memory, the system can tailor its responses to the individual user, improving the overall user experience. The \texttt{recall\_memories\_templated} function is the cornerstone of the system's ability to recall past interactions. It operates as follows:

\begin{itemize}
    \item It accepts a user query, the number of top memories to retrieve (\texttt{k}), a metric for similarity measurement, and identifiers for the user and bot involved in the interaction.
    \item Search the database for entries that match the query, considering the specified metric and user-bot context.
    \item The function then iterates over the retrieved memories, extracting the query, answer, and timestamp from each document. This information is formatted into a string that encapsulates the query-answer pair along with the timestamp, offering a structured representation of the interaction.
    \item These formatted strings are accumulated in a list called \texttt{template}, which serves as a collection of past interactions ready for further processing or display.
    \item Finally, the \texttt{template} list is returned, containing the formatted memories that can be used to construct prompts, analyze user behavior, or enhance the context of subsequent interactions.
\end{itemize}

\begin{algorithm}
\caption{Recall Memories Templated}
\begin{algorithmic}[1]
\Procedure{RecallMemoriesTemplated}{dsl, k, metric, uid, bid}
    \State \text{Q} $\gets$ \Call{E}{query} \Comment{Embedding model}
    \State \text{dsl} $\gets$ \text{Format user and bot ID filter}
    \State \text{memories} $\gets$ \Call{FindDocumentsByVectorIndexed}{\text{dsl}, \text{Q}, \text{metric}, \text{k}}
    \State template $\gets$ \text{empty list}
    \ForAll{document, \_ in memories}
        \State query $\gets$ document['query']
        \State answer $\gets$ document['answer']
        \State timestamp $\gets$ document['timestamp']
        \State \text{template} $\gets$ query, answer, timestamp
    \EndFor
    \State \text{return template}
\EndProcedure
\end{algorithmic}
\end{algorithm}

\subsection{My Dialogue History Vectorization Method}
\label{Dialogue History Vectorization}

\subsubsection{Dialogue Model}

The dialogue is represented as a dictionary (or associative array) in the programming language Python. This data structure is chosen for its flexibility and ease of use, allowing for the storage of key-value pairs where each key is a unique identifier for a piece of information, and the value is the data itself. The specific structure of a dialogue is as follows:

\begin{itemize}
    \item \texttt{'query'}: This key stores the user's input, which is the query or statement that initiates the dialogue. It represents the user's intent or question and is crucial for understanding the context of the conversation.
    
    \item \texttt{'answer'}: Corresponding to the user's query, this key holds the response generated by the bot. The answer is a reflection of the bot's understanding and processing of the user's input and is a key component in the dialogue exchange.
    
    \item \texttt{'user\_id'}: This key identifies the user, It is used to associate multiple dialogues with the same user, enabling the tracking of individual user interactions over time.
    
    \item \texttt{'bot\_id'}: Similarly, this key identifies the bot, It helps in distinguishing between different bots, especially in systems where multiple bots are deployed to handle different types of queries or users.
    
    \item \texttt{'timestamp'}: This key records the exact time at which the dialogue occurred. The timestamp is generated using the \texttt{time.time()} function, which returns the current time in seconds since the Unix epoch. It is essential for chronological ordering of dialogues and for time-based analysis of user interactions.
\end{itemize}

\subsubsection{Algorithm}

The \texttt{MemorizeConversation} algorithm begins by preprocessing the answer string to ensure that multiple consecutive newline characters are replaced with a single newline. This normalization step helps in maintaining consistency in the data representation. Next, the query and answer are transformed into vector representations using a text encoding function. This function is assumed to be a pre-defined method that converts text into a numerical vector format suitable for machine learning or information retrieval tasks. The choice of the text encoding function can significantly impact the performance of the algorithm, as it determines how well the semantic meaning of the text is captured in the vector space. The algorithm then calculates a weighted sum of the query and answer vectors. The weights assigned to each vector are determined by the query weight and answer weight parameters, which allow for flexibility in how much importance is given to the query versus the answer in the final vector representation. This weighted sum represents a compromise between the query and the answer, capturing the essence of the conversation. A dictionary called `mem\_data` is created to store the conversation metadata, including the original query, the processed answer, user and bot identifiers, and a timestamp. This metadata is crucial for later identifying and retrieving the conversation from the database. The algorithm then creates a list of inverse indices from the keys of the `mem\_data` dictionary. These indices are used to facilitate quick searches and retrievals in the database.

\begin{algorithm}
\caption{Memorize Conversation}
\begin{algorithmic}[1]
\Procedure{MemorizeConversation}{Q, A, uid, bid, $W_Q$, $W_A$}
    \State $answer \gets \text{Replace newline characters in } answer$
    \State $V_Q \gets E(query)$ \Comment{Embedding model}
    \State $V_A \gets E(answer)$ \Comment{Embedding model}
    \State $V \gets W_Q \times V_Q + W_A \times V_A$ 
    \State $mem\_data \gets \{'Q': Q, 'A': A, 'uid': uid, 'bid': bid, 'time': \text{timestamp}\}$
    \State $inv\_indices \gets \text{list of keys in } mem\_data$
    \State \text{store memory and the corresponding unique vector index}
\EndProcedure
\end{algorithmic}
\end{algorithm}

In the semantic vector space, each text query and its corresponding answer are represented as vectors $V_Q$ and $V_A$, respectively. These vectors are embedded in a high-dimensional space where the semantic meaning of the text is captured. The weighted sum of these vectors, as shown in the equation below, is used to represent the combined semantic meaning of the query-answer pair:

\[ V = W_Q \times V_Q + W_A \times V_A \]

\noindent Here, $W_Q$ and $W_A$ are the weights assigned to the query and answer vectors, respectively. Geometrically, this operation can be visualized as follows:

\begin{itemize}
  \item The vector $V_Q$ represents the semantic content of the query.
  \item The vector $V_A$ represents the semantic content of the answer.
  \item The weights $W_Q$ and $W_A$ act as scaling factors that determine the contribution of each vector to the final combined vector $V$.
  \item When $W_Q > W_A$, the resultant vector $V$ will be closer to the direction of $V_Q$, indicating that the semantics of the query have a greater influence on the combined representation.
  \item Conversely, when $W_A > W_Q$, the resultant vector $V$ will be closer to the direction of $V_A$, indicating that the semantics of the answer have a greater influence.
  \item The angle between $V$ and either $V_Q$ or $V_A$ is smaller when the weight of that particular vector is heavier.
  \item The cosine of the angle between two vectors is a measure of their similarity. A smaller angle between $V$ and one of the vectors (say $V_Q$) implies a higher cosine similarity, indicating that the resultant vector $V$ is more similar to $V_Q$.
  \item This weighted sum approach allows us to control the balance between the query and answer semantics in the combined representation, which is crucial for applications where the importance of the query or answer may vary.
\end{itemize}

\noindent To further illustrate the geometric interpretation, consider the following diagram:

\begin{figure}[h]
    \centering
    \includegraphics[width=0.5\textwidth]{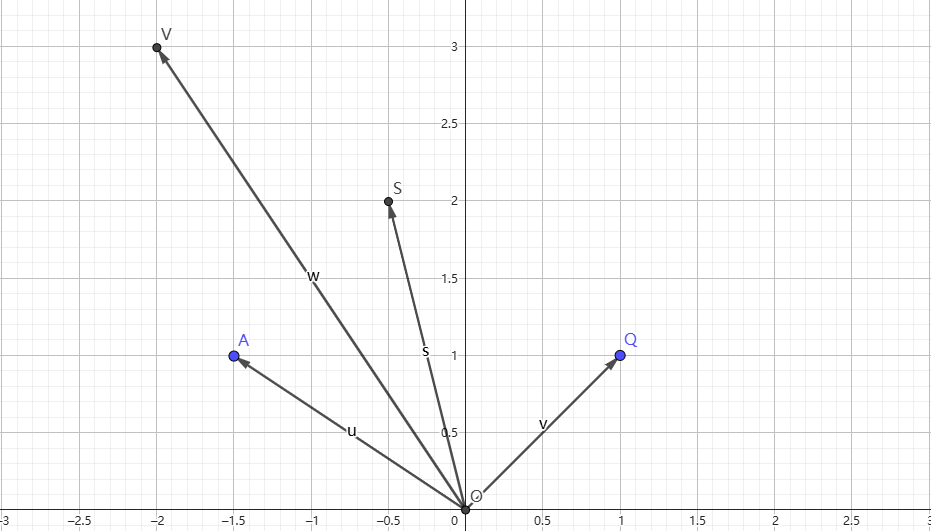}
    \caption{Geometric Interpretation}
    \label{fig:image_5}
\end{figure}

In this diagram, the vectors $V_Q$, $V_A$, and $V_S$ are represented in the semantic vector space. The weighted sum vector $V_V$ is depicted as being closer to the vector $V_A$, which is assigned a heavier weight, thereby demonstrating the influence of the weights on the final combined vector. Additionally, the angle between $V_V$ and $V_A$ is smaller compared to the angle between $V_S$ and $V_A$, illustrating the weighted sum's tendency to align more closely with the more heavily weighted vector.

\subsection{Experimental Results and Analysis}

In the dialogue records devoid of memory, as illustrated in Table \ref{tab:dwmr}, the lack of retention of past interactions becomes a pronounced impediment to the delivery of tailored and contextually aware responses. This shortcoming is particularly noticeable in the absence of any recollection regarding the nuances of human's personal life, the details of his material possessions, and the specifics of his professional endeavors. The inability to draw upon prior exchanges results in a disjointed interaction that lacks the warmth and relevance that memory can provide.

Conversely, the dialogue history presented in Table \ref{tab:mrd} showcases the profound benefits of having the capacity to remember and reference past interactions. This capability allows for a rich tapestry of conversation where details about human's life, possessions, and career are not only remembered but also woven into the fabric of subsequent discussions. The presence of a memory infuses the interaction with depth and continuity, transforming it into a more immersive and significant experience. It is this ability to recall and leverage past dialogues that endows the interaction with a sense of coherence and personalization, thereby making it more engaging and meaningful.

\begin{table}[h]
\centering
\label{tab:dwmr}
\begin{tabular}{|p{5cm}|p{5cm}|}
\hline
Human & Bot \\
\hline
Hi, my name is Test & 
Hi Test, nice to meet you. \\
\hline
I have a wife,a son and a dog. How do you like my life? & 
It sounds like you have a loving family with a wife,a son, and a dog. Your life seems fulfilling and full of joy. \\
\hline
I have a car which is cool and designed in Japan. & 
Your car sounds awesome!It must be really stylish with its Japanese design. \\
\hline
My job is to develop software in Python. & 
That's great! Python is a versatile programming language for software development. \\
\hline
What do you know about my family? &
I'm sorry, but I do not have any information about your family in my memory. \\
\hline
How do you like my taste in cars? &
I'm sorry, but I don't have the ability to perceive your taste in cars. \\
\hline
What does one need to learn if I need him(her) to help me do my job? &
To help you with your job,one would need to learn the specific tasks and requirements of your job, as well as any necessary skills or knowledge related to the job duties. \\
\hline
What do you know about me? &
I'm sorry, but I don't have any information about you. \\
\hline
\end{tabular}
\caption{Dialogue Without Memory Retrieving}
\end{table}


\begin{table}[H]
\centering
\label{tab:mrd}
\begin{tabular}{|p{5cm}|p{5cm}|}
\hline
Human & Bot \\
\hline
Hi, my name is Test & 
Hi Test, nice to meet you. How can I assist you today? \\
\hline
I have a wife,a son and a dog. How do you like my life? & 
Hi, it sounds like you have a loving family with a wife,a son, and a dog. Your life seems fulfilling and joyful. \\
\hline
I have a car which is cool and designed in Japan. & 
Hi, your car sounds cool and has a stylish design from Japan. \\
\hline
My job is to develop software in Python. & 
Hi, your job sounds interesting. How can I assist you today? \\
\hline
What do you know about my family? &
Hi, based on the information you provided, it seems like you have a loving family with a wife, a son, and a dog. Your life sounds fulfilling and joyful. \\
\hline
How do you like my taste in cars? &
Hi, based on the information you provided, it seems like you have a great taste in cars with a cool and stylish design from Japan. \\
\hline
What does one need to learn if I need him(her) to help me do my job? &
One would need to learn Python programming language to develop software. \\
\hline
What do you know about me? &
Hi, based on the information available, it seems like you have a loving family with a wife, a son, and a dog. Your life sounds fulfilling and joyful. \\
\hline
\end{tabular}
\caption{Dialogue With Memory Retrieval}
\end{table}


\section{Limitations of the Bhakti Database}

Despite the numerous advantages of the Bhakti database, it is not without limitations. The lightweight nature of Bhakti, while beneficial for small to medium-sized datasets, reveals some constraints when scaling up to larger data volumes. Here, we discuss the primary limitations of Bhakti:

\paragraph{Performance with Large Datasets} One of the most significant limitations of Bhakti is its performance with large datasets. As demonstrated in my testing on different scales of data, Bhakti's query times begin to increase sharply beyond approximately 4000 data entries. This increase indicates that while Bhakti is effective for smaller datasets, it may not be the optimal solution for applications requiring high throughput or handling very large datasets.

\paragraph{Lack of Approximate Calculation Methods} Bhakti currently focuses on precise computations rather than approximate methods like Hierarchical Navigable Small World (HNSW) \cite{Bentley1975}. This focus means that for applications where speed is critical, and approximate results are acceptable, Bhakti may not perform as well as databases designed for approximate nearest neighbor search.

\paragraph{Limited Support for Complex Queries} Bhakti's domain-specific language (DSL) for document-based pattern matching pre-filtering is a powerful feature, but it may not support all the complex query operations that more robust vector database solutions offer. This limitation could affect the flexibility of the system in handling intricate data retrieval and analysis tasks.

\paragraph{Scalability Concerns} While the Bhakti system is designed with a client-server model that allows for independent scaling of components, its scalability in a real-world, large-scale production environment has not been extensively tested. The separation of concerns between the client, server, and storage engine, while beneficial for maintenance and upgrades, may present challenges when scaling to meet the demands of a high-volume, enterprise-level application.


\section{Conclusion}

Designed with a focus on small to medium-sized datasets, Bhakti offers a light-weight and user-friendly solution that integrates seamlessly with Python3, making it accessible for a wide range of applications. The system's support for various similarity calculation methods and its domain-specific language for document-based pattern matching pre-filtering enhance its flexibility and utility in data management tasks.

My innovative memory-enhanced dialogue solution sets it apart by assigning variable weights to question and answer pairs within dialogue history, allowing for nuanced control over the semantic significance of each segment. This feature is particularly beneficial in improving the coherence and humanness of chatbot interactions. The experimental validation of Bhakti's semantic search capabilities and its application in question-answering systems demonstrate its effectiveness in providing contextually rich and meaningful search results.

However, it is important to acknowledge the limitations of Bhakti when dealing with large datasets. The system's performance tends to decline with an increase in data volume, and it currently lacks support for approximate calculation methods, which could impact its speed in certain applications. Additionally, Bhakti's resource intensity and limited support for complex queries may restrict its suitability for high-volume transactional environments or scenarios requiring extensive data retrieval and analysis.

Looking forward, the future development of Bhakti could address these limitations by incorporating approximate calculation methods, optimizing its architecture, and enhancing its scalability features. This would broaden Bhakti's applicability and strengthen its position as a versatile tool in the realm of vector databases.

This research makes several significant contributions to the field. First, it proposes Bhakti, a lightweight and easily deployable solution specifically designed for small and medium-sized datasets. Second, it introduces support for a variety of similarity calculation methods, thereby enhancing the flexibility and applicability of the database across diverse scenarios. Third, the research achieves seamless integration of Bhakti with Python3, which significantly improves the flexibility and development efficiency of data management systems. Furthermore, the study proposes a memory-enhanced dialogue solution based on the Bhakti database. This solution enables fine-grained control over the semantic importance of each segment in dialogue history through weight assignment, thereby optimizing contextual understanding in dialogues. Lastly, the research provides a critical analysis of Bhakti's limitations in processing large datasets, clarifying its scope of application and suitability for different scenarios. These contributions collectively advance the understanding and practical implementation of database solutions in semantic search and dialogue systems, particularly for small and medium-sized datasets.

\bibliographystyle{unsrt}
\bibliography{ref}

\end{document}